\documentclass[twocolumn,preprintnumbers,superscriptaddress]{revtex4}
\usepackage{epsf}
\usepackage{graphicx}% Include figure files
\usepackage{dcolumn}% Align table columns on decimal point
\usepackage{bm}% bold math
\def\prb{Phys. Rev. B}
\def\prl{Phys. Rev. Lett.}

\def\be{\begin{equation}}
\def\ee{\end{equation}}
\def\ba{\begin{eqnarray}}
\def\ea{\end{eqnarray}}

\def\C60{A$_x$C$_{60}$}

\def\HgCu3{HgCa$_2$Cu$_3$O$_{8+y}$}
\def\HgCu4{HgBa$_2$Ca$_3$Cu$_4$O$_{10+y}$}
\def\TlCu3{Tl$_2$Ba$_2$Ca$_2$Cu$_3$O$_{10+y}$}
\def\TlCu4{Tl$_2$Ba$_2$Ca$_3$Cu$_4$O$_{12+y}$}

\def\BiCu3{Bi$_2$Sr$_2$Ca$_{2}$Cu$_3$O$_y$}

\begin{document}

%\draft

\title{ Phases intermediate between the two dimensional electron liquid
and Wigner
crystal}
\author{Boris Spivak}
\email[]{spivak@dirac.phys.washington.edu}
%\homepage[]{Your web page}
%\thanks{}
%\altaffiliation{}
\affiliation{Department of Physics, University of Washington, Seattle,
WA 98195}
\author{Steven A. Kivelson}
\email[]{stevek@physics.ucla.edu}
%\homepage[]{Your web page}
%\thanks{}
%\altaffiliation{}
\affiliation{Department of Physics, University of California, Los
  Angeles, California  90095}

\begin{abstract}

 We show that there can be no direct first order transition between  a
Fermi liquid and an
insulating electronic (Wigner) crystalline phase in
a clean
two-dimensional electron gas in a metal-oxide-semiconductor field-effect
transistor
(MOSFET);  rather, there  must
always
exist intermediate ``micro-emulsion''  phases, and an accompanying
sequence of
continuous
phase transitions.
Among the
intermediate phases which we find are a variety of electronic liquid
crystalline
phases, including
stripe-related analogues of classical smectics and nematics.
The existence of these
phases can be established  in the
neighborhood of the phase
boundaries on the basis of an {\it asymptotically exact} analysis, and
reasonable
estimates can be made concerning
the ranges of electron densities and
device geometries in which they exist.
They likely occur in clean Si MOSFETs in the range of
densities in which  an ``apparent metal to insulator
transition'' has been observed in existing experiments.
We also point out that, in analogy with the Pomaranchuk effect in
$^3$He, the
Wigner
crystalline phase has higher spin entropy than  the liquid phase,
leading to an
increasing tendency to crystallization with increasing temperature!
%We also
%mention some parallels between the intermediate electronic phases in
%MOSFETs and  in  strongly correlated materials such as
%the giant magneto-resistance manganites and the high
%temperature superconducting cuprates.

\end{abstract}


\maketitle

%\section{Introduction}

In discussions of the theory of the two dimensional electron gas (2DEG),
it  is
generally accepted that,
as a function of electron density $n$, there is a first order quantum
($T=0$) phase transition from a high density liquid\cite{caveat1}
to a low density Wigner crystalline phase\cite{caveat2}.
 This assumption is reasonable in the
case of a triangular Wigner crystal due to the presence of cubic
invariants in the Landau free energy\cite{chaikin}, and for other
lattices due to the general expectation\cite{brazovskii} that
fluctuations will always render a freezing transition first order.
The transition is
thought to occur when the dimensionless ratio $r_s\equiv [\pi
n(a_B)^2]^{-1/2}$
exceeds
a critical value\cite{cip} $r_s=r_c \sim 38$, where $a_B$ is the
effective Bohr radius in the semiconductor.  However,
this generally accepted picture is manifestly incorrect for the 2DEG in
a
metal-oxide-semiconductor field effect transistor (MOSFET), and possibly
more
generally!

Each electron in the 2DEG in a clean MOSFET
 drags along with it an  image charge in the ground-plane above.
Consequently, at small separations,  the interaction between the
electrons is
the $V(r)\sim e^2/\epsilon r$ Coulomb interaction, while for separations
larger
than the
distance to the gate,
$d$, it is the repulsive dipole-dipole interaction,
$V(r) \sim 4e^2d^2/\epsilon r^3$.  (Here $\epsilon$ is the
dielectric constant of the host semiconductor.)
In 2D systems with dipolar interactions, the following simple
argument leads to the concussion that first order phase transitions
 are forbidden:
In systems with
interactions that fall more rapidly than $1/r^2$, there
 exists a ``forbidden'' range of
densities in the neighborhood of a first order phase transition
 where macroscopic phase separation reduces the free energy of the
system.
However, when we come to compute the surface tension between  two
macroscopic phases, we find that $1/r^3$ interactions are
marginal:  for shorter range interactions,  there is a well
defined scale independent surface tension, $\sigma$,  while for
longer range interactions, $\sigma$ is scale dependent.
Specifically, for dipolar interactions, the interfacial
contribution to the free-energy  of an arbitrary macroscopic
mixture of two phases is (see, {\it e.g.} \cite{lineinta,lineintb})
\begin{equation}
F_{\sigma}=\int d{s} \sigma_0(\hat {\bf \theta})-
\frac {\sigma_1} 2\int \frac {d {\bf l} \cdot d{\bf l'}}
{\sqrt{|{\bf l-l'}|^2+d^2}}.
\label{interface}
\end{equation}
Here, the arclength integral, $ds$, runs along\cite{oriented} the
interfaces between
the two phases,
$\hat {\bf \theta}(s)$ is
the local orientation of the interface, $\sigma_0({\hat\theta})$ is
the (in general orientation dependent and by assumption positive)
short-range piece of the surface tension,
$d{\bf l}$ runs along the interfaces\cite{oriented},
$\sigma_1=2e^{2}(\Delta n)^{2}d^{2}/\epsilon$, and $\Delta n$ is
the density difference between the coexisting phases.
The second (non-local) term in Eq. \ref{interface} comes from the
long-range parts of the dipolar interaction.
One can also view it as
the leading finite size correction to the capacitance of
parallel-plate capacitors due to the fringing
fields\cite{LandauContmed}.

It is important to note that the second term
in Eq. \ref{interface} gives a negative
contribution to the effective
surface tension
which diverges
logarithmically with length; for example,  an isolated
 straight segment of
interface of length $L$ has $F_{\sigma}= L\left\{\sigma_0
-\sigma_1\log[L/2d]\right\}$.  This  implies
that there is an absolute instability of the macroscopically phase
separated state - in the regime of the phase diagram where a
classical Maxwell construction would lead to two-phase
coexistence, a state formed from a ``microemulsion'' of the two
phases (with a character and length scale to be determined),
 has lower free-energy! Thus, instead of a first order
transition between two phases, there must always be an
intermediate regime in which one or more microemulsion phase
occurs, bounded by one or more line of continuous phase
transitions.

At this point we would like to compare this situation with the
Coulomb case
(no ground plane)
where
macroscopic phase separation is forbidden. The nature of the  phases
that
result from  the ``Coulomb frustrated phase
separation\cite{kiv1,tokura}'' in what
would
otherwise have been the forbidden range of densities is an issue of
potentially
relevance in many highly correlated materials.  However, the
inhomogeneities that occur in
this situation are typically microscopic in scale, and so difficult to
distinguish
from more
familiar
charge density wave structures\cite{za}.  Moreover,  the
relevant microscopic
details
are difficult to treat with any degree of rigor.  (It is an
interesting\cite{vadim}
question, which we would like to reopen, whether there are intermediate
phases
between
the Fermi liquid and Wigner crystal phases in the 2DEG with pure Coulomb
interactions.)

The character of the microemuslion of the two coexisting phases is
determined by
minimizing $F_{\sigma}$ in Eq. \ref{interface};  the result depends on
how
anisotropic the function
$\sigma_0(\hat {\bf \theta})$ is.
The case where $\sigma_{0} (\hat{\bf \theta})$ is  independent
 of $\hat{\bf \theta}$ has been considered
in  different contexts, including lipid films ({\it e.g.} Ref.
\cite{polimer}), two dimensional uniaxial ferromagnets
({\it e.g.} \cite{Doniah}), and the
2DEG in MOSFET's \cite{trug,spivakPS}.
The resulting phase
diagram includes both stripe and bubble phases, with stripes
preferred in the center of the phase separated
region and bubbles generally thought to be slightly lower in
energy when one phase is in extreme minority.
 Current estimates\cite{lineintb}
place the difference between the dilute stripe and bubble energies at
about 6\%. In the earlier
literature, it was assumed\cite{Doniah,polimer} that there is a
direct first order transition  between   uniform stripe and
bubble
phases. This
is incorrect, even at mean-field level, since, as
we have shown, first order phase transitions
are forbidden. Thus,
a
sequence
of continuous
phase transitions (which we discuss below) must replace
the putative first order
transition \cite{spivakPS}.

In the present case, where at least one of the two coexisting phases is
crystalline,
the angular dependence of
$\sigma_0(\hat{\bf \theta})$ is not negligible, reflecting the tendency
of
crystals to facet.
Clearly, a strong angle dependence of  $\sigma_0(\hat{\bf \theta})$
tends to favor
stripe phases
(where all interfaces lie along the direction in which
$\sigma_0(\hat{\bf
\theta})$ is minimal)
relative to any form of bubble phase.

In the present paper, we characterize the phase diagram, and in
particular the
universal aspects of
the intermediate phases and phase transitions that are expected at low
or zero
temperature in an ideal MOSFET
({\it i.e.} in the absence any disorder).  We will consider explicitly
the case in
which $d$ is large compared to
the spacing between electrons, $nd^2 \gg 1$, as in this limit (as we
shall see)
fluctuation effects are
parametrically small and an appropriate mean-field theory provides a
valid zeroth-
order description of the phases.
In Sec. \ref{meanfield}, we first discuss the mean-field phase diagram,
then
in Secs. \ref{thermal} and
\ref{quantum} we discuss the effects of weak thermal and quantum
fluctuations,
respectively.  In Sec.
\ref{experiment}, we discuss some of the implications of the present
results for
the properties of real devices
(which, alas, have non-negligible disorder), and in Sec. \ref
{extensions} we
discuss some incompletely developed
ideas concerning further implications of the present line of analysis.

\section{ Mean-Field Phase Diagram}
\label{meanfield}
Two dimensionless parameters
determine the physics of the 2DEG in a MOSFET, $r_s$ (defined above)
and $a_B/d$. Let us start with a discussion of  the zero
temperature mean-field phase diagram of this system, assuming only
uniform
 states. If $nd^{2}\gg 1$, the free-energy  per
unit area
 can be represented by the
sum $f(n)=f^{(C)}+f^{(el)}$ of
 the energy
density of a capacitor $f^{(C)}=(en)^{2}/2C$ and the
internal free-energy density of the electron liquid $f^{(el)}$.
Here $C=(\epsilon d)^{-1}$ is the capacitance per unit area.
At high
 electron densities,
$r_s \ll 1$,
the kinetic energy of the electrons is much larger than their potential
energy, so the system forms a Fermi liquid.
At small densities
$r_s \gg 1$  (but still $nd^{2}\gg 1$) the
  Coulomb energy of the electrons is much larger than the kinetic
energy, so
the ground state is
 crystalline.

However, at even smaller densities when $n d^{2}\ll 1$, the
electrons interact only via dipole interactions, so the kinetic
 energy is
larger than the potential,  and the system again has a Fermi
liquid groundstate. (See discussion surrounding Eq. \ref{smalld}.)  For
$d/a_B \gg 1$, this implies that the
%** zero temperature **
phase diagram  of the system has reentrant
transitions as a function of $n$ (along the dashed-dotted trajectory in
Fig. 1)
from
a Fermi liquid phase for
$n > n_{c} \approx r_c^{- 2}(\pi a_B^2)^{-1}$  to a Wigner crystal
phase for $n_{c} > n > n_{c1} \sim (\pi d^{2})^{-1}$, to  a Fermi
liquid phase for $n_{c1}> n$.
 With decreasing
$d/a_B$, $n_{c1}$ and $n_c$ move toward each other, until
for $d<d_{c}\sim r_c a_{B}$, the Wigner-crystal phase disappears
entirely.
This is represented by the dashed line in Fig. 1.

\begin{figure}
%  \centerline{\epsfxsize=5cm \epsfbox{FETFig1.eps}}
% \centering
\epsfxsize=8cm
\centerline{\epsffile{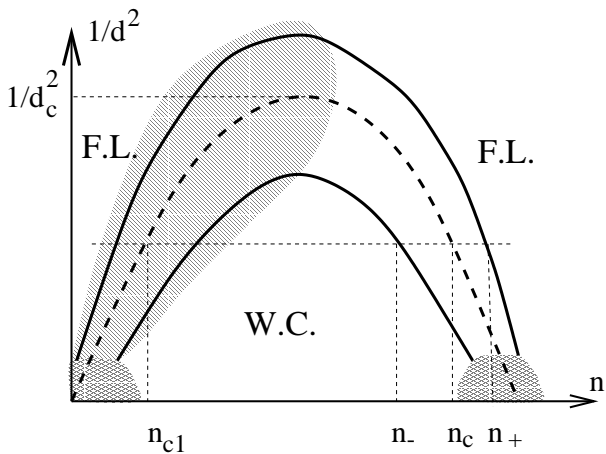}}
  \caption{The $T=0$ phase diagram of the 2DEG in an MOSFET.  The dashed
line
indicates the mean-field critical density, $n_c(d)$, where the  free
energies of the uniform Wigner crystal (W.C.) and Fermi liquid (F.L.)
phases
cross. The
solid  lines mark the boundaries of the regime of the intermediate
microemulsion (stripe or bubble)
phases.
At mean-field
level,  these solid lines are Lifshitz transitions.
They approximately coincide with the regime of macroscopic two-phase
coexistence
($n_- < n <n_+$) derived from a Maxwell construction.  The hatched area
represents
the
regime in which the
regions  of the two coexisting phases have sizes of order the electron
spacing, so
quantum
fluctuations are order 1, and hence may substantially alter the
mean-field
character of
the phases and phase transitions.  The cross-hatched areas denote the
regimes of
Coulomb frustrated phase separation where even the mean-field character
of the
phase
diagram is not known. }
%\
  \label{fig1}
\end{figure}

As a next step, we improve this phase diagram by allowing for the
possibility of
inhomogeneous
states.
There is a range of
forbidden densities about the critical density in which macroscopic
phase
separation into
regions of high and low density phase has lower free energy than the
uniform
state.

Let us briefly review the salient features of the Maxwell construction
for phase
coexistence,
as applied in the present
context.  For given average density, $n$,
 we
consider a state in which a fraction,
$x$, of the system is at a higher than average density, $n_+> n$, and a
fraction
$(1-x)$ is at a
lower than average density,
$n_-<n$, such that $xn_++ (1-x)n_-=n$.  We then minimize the total free
energy
with respect to
$n_+$ and
$n_-$.  The result of this minimization is an implicit expression for
the
densities  of the two
coexisting phases,
\be
\mu_{+} +\frac {n_+}C= \mu_-+\frac {n_-}C = \frac{[f(n_-)-f(n_+)]}
{\Delta n},
\label{Maxwell}
\ee
where $\mu_{\pm} = -\partial f^{(el)}(n_{\pm})/\partial n_{\pm}$ are the
chemical
potentials in the two phases, and $\Delta n\equiv [n_+-n_-]$.  Phase
coexistence
occurs for $n_- < n <
n_+$, where the fraction of the two phases is determined by the lever
rule,
\be
x=(n-n_-)/\Delta n.
\ee

Eq. \ref{Maxwell} is somewhat complicated, but it can be greatly
simplified
when the forbidden region is relatively small ($\Delta n \ll n_c$);  in
this case,
we can
linearize the density dependence of the free energy about the critical
density,
\be
f^{(el)}(n_{\pm}) =
f^{(el)}(n_c)-\mu_{\pm} (n_{\pm} - n_c) + \ldots
\ee
where $\ldots$ represents higher order terms in powers of  $(n_{\pm} -
n_c)$.  To
this level of
approximation,
\be
n_{\pm} = n_c \pm \frac {\Delta n} 2\  ;\
\ \ \Delta n = \frac {(\mu_--\mu_+)\epsilon}{e^2d}.
\label{npm}
\ee
The discontinuity of the chemical potential, $(\mu_--\mu_+)>0$, is
determined by
microscopic
physics, and is only small to the extent that the putative transition is
weakly
first order.
Whether or not the transition is strongly first order, for
$d$ large,
$\Delta n$  is self-consistently small.

The validity of the Maxwell construction rests on the implicit
assumption that the interface energy between the coexisting phases
is positive,  so the amount of interface is minimized.
As we have seen, in the dipolar case this assumption is invalid.
We can construct a state with lower free energy by making an
inhomogeneous mixture of the two coexisting phases to increase the
amount of interface.
To complete the mean field analysis, one should minimize Eq.
\ref{interface} with
respect to the shape
of the minority phase regions at given area of the phase - the area
being given,
to first
approximation, by the Maxwell rule.

\subsection{Stripe Phases}

To begin with, let us consider only striped phases.  This is
fully justified in the case of strong anisotropy of the surface energy.
(As
we will see letter even in the opposite case when
$\sigma(\hat{\theta})$ is isotropic, there are regions in the phase
diagram where this assumption is relevant.) The interfacial
free
 energy density for a striped phase is
easily computed from Eq. \ref{interface} to be
\begin{equation} f_{\sigma}=L^{-1}\left\{2\sigma_0-4\sigma_1 \ln [
L_+L_-/{dL} ]\right\},
\label{stripe}
\end{equation}
where $L_{\pm}$ are the widths of  the
high and low density regions, respectively, and $L=L_++L_-$ is the
period of the stripe structure.
Minimizing Eq. \ref{stripe} at fixed
areal fraction of the high density
phase,
$x\equiv L_+/L$,  we get
\begin{equation}
L_-= \frac d {x} e^{1+\gamma} \ ; \ \ \ L_+ = \frac d {(1-x)}
e^{1+\gamma}
\label{L}
\end{equation}
with $\gamma=\sigma_0/2\sigma_1$. It is important to note that as $x\to
0$, the
stripes of the high
density phase approach a finite limiting width, $L_+\to L_0=d
e^{1+\gamma}$,
although the spacing
between stripes, $L_-$, diverges in proportion to $1/x$.  Also,  because
the
minimized value of
$f_{\sigma}=-4\sigma_1/L$ is negative, the region of stability of the
striped
phase
in fact extends
somewhat beyond the edges ($n_-$ and $n_+$) of the two-phase region
derived from
the
Maxwell construction.

Finally, it is necessary to estimate the magnitude of $\gamma$; if it is
of order
1,
then $L_0\sim d$, but if $\gamma \gg 1$, then $L_0$ is exponentially
larger than
atomic lengths.
So long as the stripe phase occurs in a relatively narrow range of $n$,
we can use
Eq. \ref{npm} to
estimate $\sigma_1$, with the result that $\gamma \sim
\sigma_0e^2/\epsilon[\mu_+-
\mu_-]^2$, which is
a ratio of microscopic electronic energies.  Thus, except under special
circumstances, we expect that
$\gamma\sim 1$, and hence that $L_0\sim d$.
However, so long as $nd^2\gg 1$, the stripe widths are still large
compared to the
spacing between
electrons, which validates the macroscopic approach taken here.

In short, at mean field level, as a function of decreasing density the
system
evolves
from the Fermi liquid  phase, through intermediate stripe phases, to the
Wigner
crystal,  as
summarized in Fig. 2a:
\begin{itemize}

\item{ 1) }
 Starting in the uniform Fermi liquid phase, as the density is
varied across $n_{+}$, the system undergoes a transition to a
stripe phase, consisting of a  periodic array of far  separated
stripes of Wigner crystal, with characteristic width $L_0$.  This
transition is analogous to a Lifshitz transition, in that the
period of the ordered phase diverges at the transition
\cite{lifshitz}. Thus, the arguments\cite{brazovskii} that
fluctuations will generally drive an otherwise continuous freezing
transition
first
order
do not apply;
the continuous character
of this transition is robust.

\item{2)}  There is, of course, some coupling between the
translational motion of the crystalline order in neighboring
stripes, so at mean-field level the crystalline order will be
locked from stripe to stripe.  Consequently,
the
stripe ground-state breaks translation symmetry not only in the
direction perpendicular to the stripes, but along the stripe
direction as well.  However, near the transition, where $x\ll 1$,
the spacing between stripes is large compared to $L_0$, so this
coupling is exponentially small;  consequently, this locking can be
neglected
for all practical purposes.  Therefore, this phase
should operationally be
classified as an electron smectic\cite{nature}, in which translation
symmetry
is unbroken along the stripe direction.  (There remains the
interesting academic question of principle whether or not quantum
fluctuations are able to truly stabilize this smectic phase at
$T=0$ - this is closely related to the issue of whether ``floating
phases'' are stable in quasi-1D electronic
systems.\cite{nature,floating}).

\item{3)}  Near $x=1/2$, the stripes of Wigner crystal and the
intervening Fermi liquid are comparable in width.  As $x \to 1$,
the system is better thought of as stripes of Fermi liquid
separated by broad regions of Wigner crystal.  At some point, the
crystalline order becomes so rigid that the coupling across the
liquid stripes is no longer negligible.  At this point, the
striped state is fully crystalline, in the sense that translation
symmetry is broken in both directions, and the structure factor
contains Bragg peaks.  However, this phase is still qualitatively
distinct from the Wigner crystal. Since generally speaking the
Fermi wave vector is unrelated to the Bragg vectors of the Wigner
crystal, the liquid in the stripe ``rivers'' can still conduct
current in the stripe direction. For want of a better name, we
christen this state a striped ``conducting crystalI.'' (See
Fig.2).

\item{4)}  At $x=1$, the transition from the conducting to  Wigner
crystal
 mirrors the smectic
to Fermi liquid   transition;  it is also a Lifshitz transition at which
the
period of the
stripe order diverges.
\end{itemize}

\subsection{Bubble Phases}

So far, this analysis ignores the possibility of bubble phases.  Whether
or not
there
 is a regime in which the
lowest energy mean-field state is a bubble phase depends, as we
mentioned before,
on the
degree of anisotropy of the microscopic surface tension,
$\sigma_0(\hat{\bf
\theta})$.  It may happen,
due to the anisotropy of the Wigner crystal, that
$\sigma_0(\hat{\bf
\theta})$ is sufficiently anisotropic that bubble phases never intrude
upon the
phase
 diagram.  It is also
possible to force the issue by artificially enhancing the anisotropy of
$\sigma_0(\hat{\bf \theta})$.  This can be done by explicitly breaking
the
rotational symmetry of the 2DEG, for
instance by applying an in-plane magnetic field or  by using a
sufficiently
anisotropic surface in the
construction of the MOSFET.  In this case, no more need be said.

However, the Wigner crystal is generally thought to be triangular.
In this case the surface energy is sufficiently isotropic that for
$x$ near 0 or 1, there will be a range of $x$ in which bubble
phases have lower energy than the stripe phase; for $x$ near 0,
such a phase consists of far separated crystallites in a metallic
sea, while for $x$ near 1, it is far separated bubbles of fluid in
a Wigner crystalline host. We will call these phases Bubble Crystals I
and II,
respectively. (See Fig.2d) As $x\to 0$ or $x\to 1$, the
period of the bubble crystals diverge, leading at
mean-field level to another Lifshitz transition, much as in the
stripe case. (However, fluctuation effects are much different near these
transition in the bubble and stripe cases, as we will discuss in
the next sections.)

However, this is not the end of the story.
The stripe phase is always the lower energy one near $x=1/2$, so
if a bubble phase occurs for small $x$,
there must
be a critical value of
$x=x_c$ at which the energy of the bubble and stripe phases cross,
seemingly
 implying a first order transition.
Since we have proven in general that first order transitions are
forbidden,
 this first order transition, too,
must be replaced by a regime of intermediate phases consisting of
a mixture of bubble and stripe phases \cite{spivakPS}.  Now,
however, because of the large anisotropy of the stripe phase, the
surface tension between these two phases must be highly
anisotropic. Thus, this intermediate phase will most probably
 be of the form of
alternating mega-stripes of bubble and stripe phase regions.
These regions are shown in Fig.2d by hatched boxes.

\begin{figure}
%  \centerline{
%\centering
\epsfxsize=8cm
\centerline{\epsffile{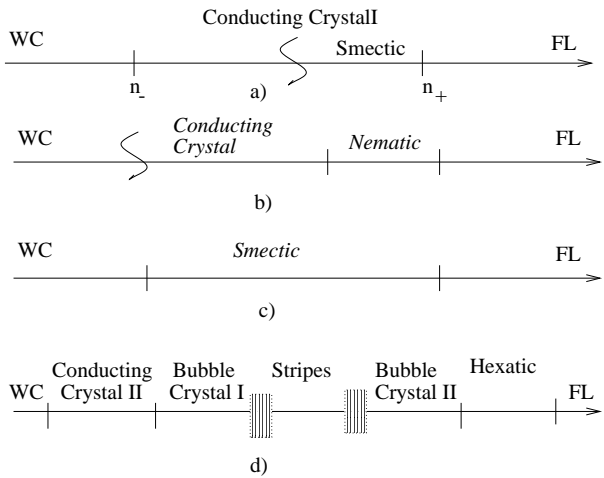}}
\caption{Schematic representation of the sequence of intermediate
states as the 2DEG
evolves from the Fermi liquid (FL) to the Wigner crystal (WC).  In
a, b, and c, we assume that $\sigma(\theta)$ is sufficiently anisotropic
that bubble phases are supressed.
  a) The mean-field phase diagram.
Under appropriate circumstances, this also represents the true
sequence of phase transitions at $T=0$.  b)  The phase diagram at
non-zero
temperature  with a rotationally invariant Hamiltonian.  c)  The
phase diagram at
non-zero temperature with a preferred axis, for instance due to an
in-field magnetic field.  Vertical lines represent phase
transitions and wavy lines crossovers. Phases with power-law order
are named in italics.  The smectic phase in a) is weakly unstable
to crystallization at mean-field level, but may be stabilized by
quantum fluctuations. d) The phase diagram including bubble
phases at $T=0$ in the presence of quantum
fluctuations. The hatched areas correspond to the sequence of
transitions involving mega-stripes of bubble and stripe phases
discussed in the text.}
%\
\label{fig2}
\end{figure}

\section{ Thermal Effects.}
\label{thermal}

\subsection{The Pomaranchuk Effect}

The most dramatic effect of finite $T$ is its effect  on the
balance between the liquid and Wigner-crystalline phases - the
fraction of the Wigner crystal phase grows as the temperature
increases \cite{spivakPS}! This phenomenon is similar to the
Pomeranchuk effect in $He^{3}$ and has the same origin: the spin
entropy of the crystal phase is substantially larger  than that of
the liquid state.  The same considerations lead, as well, to the
conclusion that the crystalline phase is preferred relative to the
liquid in the presence of an in-plane magnetic field, $H_{\|}$.

Due to the Pauli exclusion principle, an effective exchange
energy of order the
Fermi energy, $E_F^*\sim \hbar^2 n/2m^*$ quenches the spin
entropy in the liquid phase.  In contrast, the exchange\cite{exchange}
energy
 in the Wigner crystal is exponentially small, $J\propto \exp[-\alpha
\sqrt{r_s}]$ where $\alpha$ is a number of order
1. For
example an estimate made in \cite{kiv} yields $J \sim 10^{-7} Ry^*$
where $Ry^*=e^4m^*/2\hbar^2e$ is the effective Rydberg. Thus
a combination of the quantum character of the liquid
and the smallness of exchange processes in the solid imply that
the solid phase is stabilized by non-zero $T$ relative to the
liquid phase - for $n$ near $n_c$, the 2DEG freezes upon heating!
In the present context, this means that for fixed $n$, the
relative fraction of Wigner crystalline regions increases with
increasing $T$ or $H_{\|}$.
A simple estimate of the magnitude of this effect can be made for the
range of
temperatures
$J \ll T \ll
E_F^*=Ry^*/\pi(r_s)^2$ and $\hbar\mu_B H_{\|} \ll E_F^*$, where the
entropy of the liquid is negligible, as are the
 subtleties of the ground-state
magnetic structure of the Wigner crystal.
In this case,
\ba
f(n_-,T,H_{\|}) && \approx  f(n_-,0,0) \\
&& - k_BT n_- \ln\left[2\cosh(\hbar\mu_B H_{\|}/k_BT)\right] \nonumber
\label{pomaranchuk}
\ea
and $f(n_+,T,H_{\|}) \approx f(n_+,0,0)$.  The fact that temperature and
magnetic
field stabilize the Wigner crystal in
qualitatively similar fashion is one of the striking aspects of this
relation:
  For $T\gg \hbar\mu_B H_{\|}$,
$f(n_-,T,H_{\|}) - f(n_-,0,0)\approx - k_BT n_- \ln[2]$ while for
$T\ll \hbar\mu_B H_{\|}$,
$f(n_-,T,B) - f(n_-,0,0)\approx - \hbar\mu_B H_{\|} n_-$.

Of course, at high enough temperatures, all tendencies to ordered states
are
suppressed.
  This occurs above the
characteristic temperature at which the Wigner crystal melts.  In the
limit of
 very large $r_s$, this occurs
at the classical melting temperature of the Wigner crystal, which has
been
 estimated in accurate numerical
experiments\cite{sudipold} to be
\be
T_{melt}=A (e^2/\epsilon)\sqrt{\pi n} = 2A Ry^*/r_s
\ee
where $A=1/125 [1\pm 0.04]$.  However, at smaller $r_s$, where $E_F$ of
the
competing fluid phase is larger than
the putative classical melting temperature, the implied reduction of the
entropy
of the fluid state means that the
melting temperature is  set, by $T_{melt} \propto E_F$.  Far from the
Lifshitz
points, the melting temperatures of
the various microemulsion phases are determined by these same
considerations, and
are of similar magnitude.  Here, the fraction of the system that is
crystalline is a non-monotonic function of $T$,
first increasing and then dropping to zero at $T_{melt}$.  Near the
Lifshitz points, more delicate considerations
determine the melting point.

\subsection{Thermal fluctuations in the stripe phases}
Let us now consider the role of thermal
fluctuations on the
stripe
phases.  We distinguish two cases:  1)  If
the Hamiltonian is rotationally invariant, then the smectic phase is
unstable
at any non-zero temperature to the proliferation of dislocations.  Thus,

the mean-field smectic phase is replaced by a nematic phase, which, in
keeping with the Mermin-Wagner theorem, does not actually break
rotational symmetry, but rather has power-law orientational order.  A
free dislocation has a logarithmically divergent energy in both the
Wigner and conducting crystal phases, so they are robust against
thermal
fluctuations at low temperatures, although with power-law rather than
long-range crystalline order.  The resulting phase diagram is shown
schematically in Fig. 2b.  2)  If, however, the Hamiltonian has a
preferred axis, for instance if we consider the 2DEG in the presence of
an in-plane magnetic field, the effects of low temperature thermal
fluctuations are much less severe.  Here, the smectic and both
crystalline phases remain well defined at non-zero $T$, although again
with power
law
spatial correlations rather than with true long-range order, as shown
schematically in Fig. 2c.

Because first order transitions are forbidden, the transition between
the
isotropic fluid and the nematic phase must be of the
Beresinskii-Kosterlitz-Thouless (BKT) type.
Near the mean field Lifshitz
point we can estimate this transition temperature as
follows:  The distance between stripes is large
 so  the stripes of minority phase
 evaporate when the energy to break off a piece, $\sim \sigma_0L_0$, is
less than the the configurational entropy of a state where
 rare droplets of the minority phase
 are distributed randomly.
Equating these two free energies leads to the estimate
\begin{equation}
 T_{c} |\ln [x(1-x)]| \sim \sigma_0 L_0.
\label{tc}
\end{equation}

In the presence of an in-plane magnetic field, there is no sharply
defined nematic
phase, since rotational
symmetry is explicitly broken.  However, by the same token, free
dislocations in
the smectic state have a
logarithmically divergent energy, so a power-law smectic phase exists at
non-zero
$T$.  With increasing
temperature, the smectic to liquid phase transition is also of the
BKT type.
Indeed, so long as the symmetry breaking term in the Hamiltonian is
small, the
transition temperature  is roughly
the same as in Eq. \ref{tc}, above.

\subsection{Thermal fluctuations in the bubble phases}

We now consider the effect of thermal fluctuations on bubble phases.
As can be seen from Eq. 1, the interaction energy
between far separated bubbles
decreases at large $r_{0}$ as
\be
V_{bubble} \sim \sigma_1 L_0^4/ r_0^3,
\label{Vbubble}
\ee
where $L_0$ and $r_0$ are the radius of and
distance between bubbles.
  Thus, where the bubbles are far separated,
because of the screening by the ground-plane  the BKT melting
temperature will tend to rapidly to zero, $T_{BKT} \propto
[(1-x)x]^{3/2}$,
 as the spacing between bubbles
increases. The result is that, near the mean filed Lifshitz point
the bubble phase is always melted by the thermal fluctuations. On
the other hand, at smaller $r_{0}$ the bubble phase survives
thermal fluctuations in the usual sense  that the correlations of
bubble positions exhibit power-law decay.

The nature of the transition between the bubble phase and the
uniform phase is not, presently, settled. Of course, a direct
first order transition is forbidden.  One possibility is that
there is a sequence of two BKT transitions, as in the
Halperin-Nelson theory\cite{HalperinNelson} of melting, with an
intermediate hexatic phase. Alternatively, there may be a further
set of hierarchical microemulsion phases.

\section{Quantum Fluctuations }
\label{quantum}
\subsection{Stripe Phases}
So long as $nL_0^2\gg 1$
($nd^2\gg 1$), the stripes are many electrons wide, so quantum
fluctuations of their positions are intrinsically  small; $1/nL_0^2$
 is a small parameter in the problem, which
permits an asymptotically exact treatment of quantum fluctuation
effects.
  At zero temperature, the
conducting-crystal phase is clearly stable in the presence of small
 quantum fluctuations, although, as
mentioned previously, the jury is still out on whether the smectic
phase is unstable to crystallization\cite{nature,floating}.
Only where the stripe width is of order of the interelectron distance (
{\it
i.e.},
when $nL_0^2\sim 1$),
 quantum fluctuations  become very significant.
This applies
to the hatched region in Fig. \ref{interface}, where the
quantum properties of the system
are still
uncertain.

The quantum nature of the system near the Lifshitz points is
determined by the quantum nature of the interface between the
crystal and the liquid -- a problem which itself is still
unsolved. This interface may be quantum rough or quantum smooth.
If it is smooth, the Lifshitz transition from the uniform fluid
to smectic phase is not fundamentally affected by quantum
fluctuations, provided the width of the stripes is large enough.
However,
if an isolated interface is rough, the stripe order in the vicinity of
the mean-
field
Lifshitz point is quantum melted; in this case,
for the rotationally invariant system, the proscription against first
order transitions implies that there must be an intermediate zero
temperature nematic phase between the isotropic and the stripe
 ordered phases.  It was
recently shown\cite{vadim} that a nematic Fermi fluid  is
 necessarily a non-Fermi liquid in the
sense that quasiparticles are not well defined elementary excitations.
 We believe that, depending on microscopic details and on the the value
of
$nd^2$, both scenarios are possible.

It is worth mentioning why the quantum nature of the
crystal-liquid interface is so subtle. Consider the motion of a
step in the interface. Quantum-mechanically, an isolated step
might be expected to propagate along the interface \cite{andreev}.
Because the steps interact by a short-range dipolar interaction,
the steps should then form a delocalized 1D quantum liquid along
the interface.
However, because of
the density mismatch between the solid and liquid, the situation
is more complicated. Motion of the step  requires a flux of
 mass into the liquid of a magnitude proportional to the density
difference between the solid and liquid and to the step's
velocity. In a Fermi liquid this flux of mass
is carried by
quasiparticles, making the step motion highly dissipative. Thus,
characterizing the interface involves interesting, but as far as
we know unsolved issues in dissipative quantum mechanics.

\subsection{Bubble Phases}

In
contrast to stripe phases, quantum fluctuations always melt the
bubble phases when the bubbles are sufficiently dilute.  To see  this,
we can
estimate the characteristic potential energy of a bubble crystal
as in Eq. \ref{Vbubble}, and can make a corresponding dimensional
estimate
of the   bubble kinetic energy
$K_{bubble}\sim \hbar^{2}/r^{2}_{0}m^{*}$ where $m^*$ is the bubble
effective mass. Therefore the
ratio of these energies is
\begin{equation}
\frac{V_{bubble}}{K_{bubble}}\sim \left(\frac {\sigma_1 L_0^4  m^*}
{\hbar^2}\right)\frac{1}{r_{0}}
\label{smalld}
\end{equation}
 vanishes as $r_{0} \rightarrow \infty$.
This analysis fleshes out the same argument mentioned in the
introduction
 that leads to the conclusion that   there is no stable Wigner crystal
phase at small
$d$.  However, whereas in that case, the proportionality constant
is $a_B^*$, in the present case the same constant is
parametrically large, both due to the explicit factors of $L_0$
and due to the fact that $m^*$ increases with increasing $L_0$ (in
a way that depends on whether the interface is quantum rough or
smooth).  The result is that, for large $nd^2$, the regime in
which the bubble crystal is quantum melted is extremely small.
However, if $nd^2\sim 1$ quantum melting is a significant phenomenon.

The character of the bubble liquid phase is different depending on
the character of the minority phase. When the majority phase is
Wigner-crystalline with  dilute inclusions of liquid, the melting
of the bubble crystal results in a type of ``conducting
crystal''\cite{spivakPS}. In this state, crystalline long-range
order coexists with fluid-like conductivity, but in this case the
conductivity  is associated with the motion of the bubbles
themselves.  Phenomenologically this state is similar to the
``supersolid'' phase which has been
discussed\cite{andreev1} in the context of $He^{4}$.
In both cases the number of electrons (or
aroms) is not equal to the number of the crystalline sites.
\cite{comparisonAndr}. The difference is that unlike the case of
$He^{4}$ where vacancies are bosons, in our case the statistics of
the droplets is not known, and hence the liquid state may
not be a superfluid. Therefore, we refer to this state as a
"Conducting Crystal II" in Fig.2d to distinguish it
from the highly anisotropic conducting crystal (See Fig.2a,d)
which originates from the existence of stripes.

 When the Fermi liquid is the majority
phase, with a fluid of ``icebergs'' floating in it, no spatial
symmetries need be broken.  However, elementary excitation
spectrum is likely to be different from that of a conventional
Fermi liquid.

Since the majority phase already brakes rotational symmetry, the
bubble crystallization transition which transforms the system from
the conducting crystal to the insulating bubble crystal phase can
be a simple continuous transition.  However, the freezing of the
icebergs into a triangular crystal of Wigner-crystalline bubbles
is more problematic. As with the thermal transition, there may be
a two-step freezing transition, with an intermediate quantum
hexatic phase\cite{vadim}, or another hierarchy of microemulsion
phases. The sequence of the phases at $T=0$ is shown in
Fig.2d.

\section{ Experimental Consequences}
\label{experiment}
Obviously, there are many
experimental consequences of the existence of  intermediate
phases, of which we here  list only a few.  It should be kept in
mind that macroscopic spatial symmetry breaking, the sort which
precisely characterizes the various phases we have discussed, does
not truly occur in 2D in the presence of quenched disorder.  This
complicates the actual observation of various phenomena.

The majority of industrially produced Si MOSFET's have   gates
relatively close to the 2DEG, $d\ll d_{c}$, so the electron liquid
is  weakly interacting at all $n$.  However, a small number of
high mobility Si MOSFET's (For a review, see \cite{rmp}.)
and p-type of GaAs double layers  \cite{Pillarisetty} with
large $d \sim
1000\AA$ have been studied in the past few years, and found to
exhibit transport anomalies that have been interpreted as evidence
for an unexpected metal-insulator transition.  While these devices
certainly
are not ideal, in the sense that they have non-zero quenched disorder,
we would like to propose that a natural explanation of these  phenomena
is that
they reflect the existence in the zero disorder limit of the electronic
microemulsion phases identified
in the present paper.

One robust consequence of two-phase coexistence is that the
conductivity is a decreasing function of the volume fraction of
Wigner crystal.  This volume fraction, in turn, is strongly
temperature and magnetic field dependent due to the Pomeranchuk
effect, as explained above.
As a
result, the fraction of  crystal grows with temperature and magnetic
field, leading to a
corresponding increase of the resistivity.
As has been pointed out
previously\cite{spivakPS}, this basic physics may underly the
transport anomalies  observed in large $d$ Si
MOSFETs.  In particular, it offers a candidate explanation of the
anomalous
metallic ($d\rho/dT > 0$) temperature dependence and  large positive
magneto-resistance observed in these systems despite the fact that the
resistivity, itself, exceeds the Ioffe-Regel limit.  (Ideally, one might
want to
explore the
scaling relation between the temperature and magnetic field dependence
of the
resistivity implied
by the thermodynamic relation in Eq. \ref{pomaranchuk}.)

Of course, each new phase
has different patterns of spatial
symmetry
breaking, and hence has new collective modes and modified
hydrodynamics.  Even
when the
effects of quenched disorder or thermal fluctuations restore the
symmetry at
macroscopic
distances, the existence of these collective modes can have readily
detectable
consequences for the dynamical responses of the system.  Small explicit
symmetry
breaking
fields can be used to overcome the destructive
effects of quenched disorder and reveal the true tendency to symmetry
breaking.
For
instance, an in-plane magnetic field explicitly breaks  rotational
symmetry;
where
some form of stripe or nematic phase exists in the absence of  quenched
disorder, the small symmetry breaking produced by such a field can give
rise to a
large
resistivity anisotropy, as has been seen for the analogous states in
quantum Hall
devices\cite{eisenstein}.

\section{ Extensions}
\label{extensions}

We end with some   speculative observations
concerning  intermediate phases of the 2DEG.

{\it Spin physics:}  Other than the Pomeranchuk effect,
we have largely ignored the physics of the electron spins.
The
exchange interactions in the Wigner crystal phase are generally found to
be very
small\cite{exchange}, and so are only important at very low
temperatures.  At
$T=0$,
however, the fact that the magnetic Hamiltonian is highly frustrated and
may have
important multispin ring exchange interactions, can lead to a variety of
possible
magnetic phases, and this complexity could be inherited, to some degree,
by the
intermediate phases discussed here.  Moreover, at a liquid-crystalline
interface,
the
quantum dynamics of the interface itself (mentioned above) can produce
effective
exchange
interactions, likely with much larger energy scales than in the bulk
Wigner
crystal.
There is thus the very real possibility that the magnetic structure of
the
interfaces is very rich, and characterized by substantial energy scales.

{\it Superconductivity}:  The parallels between the 2DEG in a MOSFET and
Coulomb
frustrated phase separation in a doped Mott insulator naturally lead one
to {\it
speculate} concerning the possibility of superconductivity in the
present system.
 In the
bubble related conducting crystal phase, each bubble has a fixed number
of electrons;  when that number is even, the bubbles are likely
bosonic and a supersolid phase with low superfluid
density is possible \cite{spivakPS,andreev1}.   In the hatched region of
the
phase
diagram, where quantum effects are most severe, a more robust
mechanism is possible, based on the ``spin-gap proximity
effect\cite{zachar}'': Small clusters of Wigner crystal (be they
stripe or bubble like) will often have a spin-gap.  (Near the
cluster edge,
 this gap may be larger than in the bulk.)  Where this gap is large
enough, it
suppresses
single-particle exchange  between the crystal clusters and the
surrounding Fermi
fluid, but pair-exchange  is still permitted. When this dominates,
it induces global superconductivity by a process analogous to the
conventional
proximity
effect.

{\it Double Layers}:  In a
double layer system, with two nearby 2DEG's, the two layers screen each
other in much the same way as the metal layer screens the 2DEG in a
MOSFET.
However, here
the types of phases, and the available experiments are still richer.
One
particularly
interesting point is that the conductivity measured in drag can explore
the nature
of the
interlayer screening.  The presence of a crystalline component of an
electron
fluid has
the potential to greatly increase the drag conductivity relative to a
Fermi
liquid;  in
particular, whereas  the drag conductivity vanishes as $T\to 0$ in a
Fermi liquid,
we
believe it can approach a non-zero constant in some of the intermediate
phases we
have
explored.

{\it Other Applications:}  The present ideas are pretty clearly
applicable in a
host of
additional physical contexts.  What is needed is short-range tendency to
phase
separation,
{\it i.e.} a concave local free energy, opposed by dipolar forces.
Under
appropriate
circumstances, this situation may pertain in the 2DEG at higher
densities, $r_s <
r_c$,
and it certainly applies in various regimes to the physics of lipid
films and
planar
ferromagnets.

{\bf Acknowledgements:}
 This work was supported in part by the National Science
Foundation under Contracts No. DMR-01-10329 (SAK) and DMR-0228104 (BS).


\begin{thebibliography}{30}

\bibitem{caveat1}  We will refer to the high density liquid phase as a
Fermi
liquid, although
it may be that it is unstable at very low temperature to the formation
of
a high
angular momentum superconducting phase.  Nearer to the critical $r_s$
there is
also
the possibility that various other types of non-Fermi liquid states,
distinct from
those
discussed here, could occur.  None of these subtleties are relevant for
the
present discussion.

\bibitem{caveat2}  All crystalline phases with an integer number of
electrons per unit cell will here be called a ``Wigner crystal''
although this may
actually refer
to a host of distinct phases, which differ both in crystal\cite{bhatt}
and
magnetic\cite{kiv} structure.

\bibitem{chaikin} P.~M.~Chaikin, T.~C.~Lubensky, ``Principles of
condensed matter
physics'',
Cambridge University Press, 1995.

\bibitem{lineinta}     S.~Marchenko, Sov. Phys. JETP, {\bf 63}, 1315,
(1986);  S.~A.~Langer, R.~E.~Goldstein, and D.~P.~Jackson, Phys. Rev.
{\bf
A 46}, 4894, (1992)
\bibitem{lineintb}  Kwok-On and D.~Vanderbilt, Phys. Rev. {\bf B
52}, 2177, (1995).

\bibitem{brazovskii} S.~A.~Brazovskii, Sov. Phys. JETP., {\bf 41},
85, (1975).

\bibitem{cip} B.~Tanatar and D.M.~Ceperley, Phys. Rev.~B {\bf
39}, 5005
(1989).

\bibitem{oriented}  In a case where the area of the minority phase
 is not simply connected,
the integrals $d{\bf l}$ in Eq. \ref{interface} should be taken in the
clockwise
direction. The $d^{2}$ in the denominator of Eq. \ref{interface} has
been
introduced
to cut off the logarithmic divergence at short distances.

\bibitem{LandauContmed}  L.~D.~Landau and
E.~M.~Lifshitz, ``Electrodynamics of continuous media''
(New York, Pergamon, 1984).

\bibitem{kiv1} V.~J.~Emery and S.~A.~Kivelson, Physica (Amsterdam) {\bf
209}, 597,
(1993).  U.~L{\o}w {\it et al}, \prl {\bf 72}, 1918 (1994).

\bibitem{tokura}  For a recent perspective, see Y. Tokura, Physics
Today, July
2003.

\bibitem{polimer} M.~Seul and D.~Andelman, Science, {\bf 267}, 477,
(1995).

\bibitem{Doniah} T.~Garel and S.~Doniah, Phys.Rev. B, {\bf 26}, 325,
(1982).

\bibitem{spivakPS} B.~Spivak Phys.Rev.{\bf B 67}, 125205 (2003)

\bibitem{trug} S.~A.~Kivelson and S.~Trugman, Phys.Rev. B. {\bf 33},
3629, (1986).

\bibitem{bhatt}  X.~Wan and R.~N.~Bhatt, \prb {\bf 65}, 233209 (2002).

\bibitem{nature} S.~A.~Kivelson, E.~Fradkin, and V.~J.~Emery, Nature
{\bf 393},
550
(1998).

\bibitem{vadim} V.~Oganesyan, S.~A.~Kivelson, and E.~Fradkin, Phys.Rev.
B. {\bf 64}, 195109, (2001).

\bibitem{za} J.~Zaanen and O.~Gunnarson, Phys.Rev.B. {\bf 40}, 7391,
(1989).

\bibitem{kiv} S.~Chakravarty {\it et al},
%, S.~Kivelson, C.~Nayak, K.~Voelker,
Phylos. Mag. {\bf 79}, 859 (1999).
\label{sudip}

\bibitem{sudipold}  R.C.Gann, S.Chakravarty, and G.V.Chester, \prb {\bf
20}, 326
(1979).

\bibitem{floating}  V.~J.Emery {\it et al}, \prl {\bf 85}, 2160 (2000)
and
A.~Vishwanath and D.~Carpentier, \prl {\bf 86}, 676 (2001) .

\bibitem{lifshitz}  We would like to mention a difference with the
conventional Lifshitz transition. In the present case there are two
spacial scales
$L_{\pm}$, one of which diverges at the transition, while the density
difference
$\Delta n$ remains finite at the transition point.


\bibitem{eisenstein}
%For a recent review, see M.M. Fogler, Int. J. Mod. Phys. B {\bf 16},
%2924 (2002), and for
For a recent experimental investigation of the effect of
explicit symmetry breaking fields in quantum Hall nematics, see
K.~B.~Cooper
%, M. P. Lilly, J. P. Eisenstein, L. N. Pfeiffer, and K. W. West
Phys. Rev. B {\bf 65}, 241313 (2002)

\bibitem{exchange}  For estimates of the exchange energy in the Wigner
crystal,
see
M.~Roger, \prb {\bf 30}, 6432 (1984) and Ref. \cite{kiv}.

\bibitem{zachar}  V.~J.~Emery, S.~A.~Kivelson, and O.~Zachar, \prb {\bf
56}, 6120
(1997) and
{\bf 59} 15641(1999).

\bibitem{andreev} A.~F.~Andreev, A.~Yu.~Parshin, Sov.Phys. JETP {\bf
48}, 763, (1978).

\bibitem{andreev1} A.~F.~Andreev, I.~M.~Lifshitz,
Sov.Phys. JETP, {\bf 29} 1107, (1969).
\bibitem{HalperinNelson} D.N. Nelson, B.I. Halperin, Phys.Rev.
{\bf B 19}, 2457, (1979).
\bibitem{rmp} For a recent review, see E. Abrahams,
S. V. Kravchenko,
M.P. Sarachik, Rev. Mod. Phys. {\bf 73}, 251, (2001).
\bibitem{comparisonAndr} Existence of a supersolid phase
in $He$ crystals has not been proven. Existence of vacancies in
the ground state in this case is of a quantum mechanical nature
while on our case the existence of the liquid droplets embedded
into the liquid is a consequence of the electrostatics.
\bibitem{Pillarisetty} R. Pillarisetty at all; Phys.Rev.Lett. {\bf 90},
226801,
2003.
\end{thebibliography}
\end{document}